\documentclass{article}

\usepackage{arxiv}

\usepackage{amsbsy}
\usepackage{graphicx}
\usepackage{amsmath}
\usepackage{multirow}
\usepackage{comment}
\usepackage{float}
\usepackage[utf8]{inputenc} 
\usepackage[T1]{fontenc}    
\usepackage{hyperref}       
\usepackage{url}            
\usepackage{booktabs}       
\usepackage{amsfonts}       
\usepackage{nicefrac}       
\usepackage{microtype}      
\usepackage{lipsum}
\usepackage{natbib}
\usepackage{ragged2e}

\newcommand{\matX}{\mathbf{X}}
\newcommand{\matW}{\mathbf{W}}
\newcommand{\matA}{\mathbf{A}}
\newcommand{\matS}{\mathbf{S}}
\newcommand{\matH}{\mathbf{H}}

\newcommand{\matC}{\mathbf{C}}
\newcommand{\vecc}{\mathbf{c}}

\usepackage[utf8]{inputenc}
\usepackage[english]{babel}

\usepackage{amsthm}

\usepackage{appendix}

\theoremstyle{plain}
\newtheorem{theorem}{Theorem}[section]

\newtheorem{lemma}[theorem]{Lemma}

\theoremstyle{definition}

\theoremstyle{remark}

\DeclareMathOperator*{\argmin}{arg\,min}

\title{Nonnegative Matrix Factorization with Group and Basis Restrictions}

\author{
  Phillip Shreeves\\
  University of British Columbia - Okanagan\\
  Department of Statistics\\
  Kelowna, British Columbia\\
  \texttt{shreeves@alumni.ubc.ca} \\
   \And
 Jeffrey L. Andrews \\
  University of British Columbia - Okanagan\\
  Department of Statistics\\
  Kelowna, British Columbia\\
  \texttt{jeff.andrews@ubc.ca} \\
   \AND
  Xinchen Deng\\
  University of British Columbia - Okanagan\\
  Department of Physics\\
  Kelowna, British Columbia\\
  \And
  Ramie Ali-Adeeb\\
  University of British Columbia - Okanagan\\
  Department of Physics\\
  Kelowna, British Columbia\\
  \And
  Andrew Jirasek\\
  University of British Columbia - Okanagan\\
  Department of Physics\\
  Kelowna, British Columbia\\
  \texttt{andrew.jirasek@ubc.ca}\\
}

\begin{document}
\maketitle

\begin{abstract}
Nonnegative matrix factorization (NMF) is a popular method used to reduce dimensionality in data sets whose elements are nonnegative. It does so by decomposing the data set of interest, \textbf{X}, into two lower rank nonnegative matrices multiplied together ($\mathbf{X} \approx \mathbf{WH}$). These two matrices can be described as the latent factors, represented in the rows of \textbf{H}, and the scores of the observations on these factors that are found in the rows of \textbf{W}. This paper provides an extension of this method which allows one to specify prior knowledge of the data, including both group information and possible underlying factors. This is done by further decomposing the matrix, \textbf{H}, into matrices \textbf{A} and \textbf{S} multiplied together. These matrices represent an 'auxiliary' matrix and a semi-constrained factor matrix respectively. This method and its updating criterion are proposed, followed by its application on both simulated and real world examples.

\end{abstract}


\pagebreak

\section{Introduction} \label{intro}
Dimensionality reduction is an important tool in fighting the \textit{curse of dimensionality} \citep{bellman1956dynamic}. This is typically done by mapping the data to a lower dimensional subspace in some way, while also retaining as much information as possible from the original data set. There is a large number of techniques that are able to perform dimensionality reduction in this manner, including principal component analysis \citep[PCA,][]{jolliffe2011principal}, vector quantization \citep[VQ,][]{gray1990vector}, factor analysis \citep{harman1960modern}, and nonnegative matrix factorization \citep[NMF,][]{lee1999learning}. All of these techniques follow two key properties of dimensionality reduction specified by \cite{wang2012nonnegative}, which includes reducing the total number of dimensions in the data and recovering factors in the data that can be interpreted effectively.

NMF differs from the other methods specified above primarily due to the fact that it has the added constraint of nonnegativity throughout its matrices. It also benefits as the recovered components represent partitioned regions of the observations, as opposed to reduced amounts of an entire observation, which can be beneficial in many applications. For example, \cite{lee1999learning} first introduce NMF by applying it to a data set containing facial images. NMF is able to reduce the dimensionality by identifying components such as the nose, eyes, and mouth. PCA, on the other hand, identified entire facial structures as components, making it difficult to envision the observations as a linear combination of parts. 
We propose a restricted algorithm that allows for one to specify expected factors and groupings as added constraints to the model, leading to a more flexible modelling paradigm for the user.

The proposed algorithm is particularly useful in the field of Raman spectroscopy (RS), an optical interrogation method whereby vibrational modes of constituent molecules are identified through inelastic light scattering. Raman spectra can provide detailed information on a range of molecular constituents within a single sample acquisition \citep{butler2016using,feng2017raman,pence2016clinical}. Relative changes in peak areas and/or positions can be used to identify altered molecular dynamics within a system undergoing a given perturbation. Typical dimensionality reduction practice in the field of RS is to use principal component analysis \citep{pieters2013raman,harder2015raman,matthews2015radiation, koch2017raman}. However, this is not an ideal model as the constituent underlying spectra should not be permitted to take on negative values. Furthermore, it is inappropriate to make the assumption that all constituents, which in this case are represented through the principal components, are uncorrelated. This is because multiple Raman bands typically originate from a single chemical species. Therefore, one could expect these bands to be correlated in their response to a given perturbation. Thus other dimensionality reduction techniques, such as NMF, are more appropriate for application to these data sets. In the following sections, the NMF model is extended to improve overall interpretation and allow for suspected chemical constituent spectra to be specified during model fitting. Other models are also applied to this data in order to further expand knowledge in the field of RS.


\section{Background}\label{background}
\subsection{Nonnegative Matrix Factorization}\label{nmf}
Nonnegative matrix factorization was originally developed by \cite{lee1999learning} as a method of finding the major underlying factors of a model and seeing how each individual observation is composed of the factors. The factors are found by mapping the observations to a latent factor space that accounts for the vast majority of variance in the model. Each observation is then written as a linear combination of the discovered factors. These factors and coefficients of the linear combinations are found by decomposing the nonnegative data of interest $\matX \in \mathbb{R}^{n \times p}_{\geq 0}$ (where $n$ is the number of observations and $p$ is the number of variables) into two lower rank nonnegative matrices $\matW \in \mathbb{R}^{n \times q}_{\geq 0}$ and $\matH \in \mathbb{R}^{q \times p}_{\geq 0}$ such that
\begin{equation} \label{maineq}
    \matX \approx \matW\matH
\end{equation}
where with $q$ underlying factors. We can describe the span of the columns of $\matH$ as the $q$-dimensional subspace and $\matW$ as the scores for each observation on that subspace. Thus, each observation $\mathbf{x}_i$ can be viewed as an additive linear combination of the features contained in the rows of $\matH$ that are weighted by the scores contained in the columns of $\matW$. This equation is an approximation of the assumed true decomposition of $\matX$
\begin{equation} \label{eqwithnoise}
    \matX = \matW\matH + \mathbf{\varepsilon}
\end{equation}
with $\varepsilon \in \mathbb{R}^{n \times p}$ represents the residual or noise matrix of the model. This noise $\mathbf{\varepsilon} = \matX - \matW\matH$, is a key component in the model as its assumed probability distribution assists in determining what objective function the data needs to be optimized with respect to \citep{wang2012nonnegative}. For example, assuming Gaussian distributed noise, the Frobenius norm 
\begin{equation} \label{froobj}
    D_F(\matX\mid \matW\matH) = \frac{1}{2}\mid\mid \matX - \matW\matH \mid\mid^2_F = \frac{1}{2}\sum_{ij}(x_{ij}-[\matW\matH]_{ij})^2
\end{equation} 
is the function that is to be optimized \citep{lee2001algorithms,wang2012nonnegative}. Another function commonly used is the generalized Kullback-Leibler divergence (GKLD) which is used when assuming Poisson distributed noise\citep{lee2001algorithms,wang2012nonnegative}. In this paper, Gaussian noise will be assumed as the simulated data sets were created with this noise and preliminary diagnostics regarding real data sets for complex Raman spectra such as those in Section~\ref{raman} suggest this may be a reasonable assumption.

\cite{lee2001algorithms} describe that NMF can be formulated by minimizing the function $\mid\mid \matX - \matW\matH\mid\mid^2$ and while this function is convex in $\matW$ or $\matH$ individually, they are unfortunately not convex in both variables together. Because of this, finding global minima is not guaranteed and algorithms which find local minima are predominantly considered at the moment. The prototypical algorithm for finding these matrices is an alternating multiplicative updating process that uses the following updates:
\begin{equation}
    h_{ij} \leftarrow h_{ij}\frac{(\matW^T\matX)_{ij}}{(\matW^T\matW\matH)_{ij}}, \ \ \ \ \ \ \ \ \ \ \ \ \ \ \ \ w_{ij} \leftarrow w_{ij}\frac{(\matX\matH^T)_{ij}}{(\matW\matH\matH^T)_{ij}}.
\end{equation}
This updating process is considered slow by most accounts and scales dramatically in elapsed time when requiring both the rank and dimension of $\matX$ to increase \citep{lee2001algorithms, wang2012nonnegative}. Suggestions to increase the performance of the algorithm include applying gradient descent algorithms \citep{lin2007projected} or use of the conjugate gradient \citep{zdunek2007nonnegative}. While NMF is advantageous due to it's added interpretability in comparison to PCA, it does fall victim to slower computation speed \citep{devarajan2008nonnegative}. Aside from the assumption of nonnegativity, \cite{lee1999learning} also note that no further assumptions are made about the statistical independencies of the underlying factors.

\subsubsection{Extensions of Nonnegative Matrix Factorization}\label{prevcon}
Various authors have considered constraints to address questions concerning clustering and classification with the NMF algorithm. \cite{ding2005equivalence} suggest that when the Frobenius norm is used as the objective function, the standard NMF algorithm bears resemblance to a relaxed form of K-means clustering \citep{hartigan1979algorithm}  where one factor matrix contains the centroids of the clusters and the other contains indicators of cluster memberships. This is then expanded upon by \cite{li2007solving} in a semi-supervised clustering manner, where limited knowledge of cluster membership is present. \cite{wagstaff2001constrained} specify how semi-supervised clustering is an algorithm performed with respect to two separate constraints. These two constraints are that there is the existence of must-link observations where the observations must be grouped into the same cluster and that there also exists cannot-link observations that are required to be in separate clusters.

When this knowledge is limited to a number of observations $l<n$, a semi-supervised clustering algorithm is needed. \cite{liu2011constrained} address these constraints by introducing the decomposition of the feature scores matrix $\matW$ into two lower rank matrices, consisting of a constrained grouping matrix as well as an `auxiliary matrix'.  Here, the grouping matrix uses the \textit{a priori} label information and stores it in the form of an indicator matrix $\matC=\left[\begin{smallmatrix}\vecc_1 & \vecc_2 & \ldots & \vecc_g\end{smallmatrix}\right]$ where $g$ is the total number of classes. This matrix is then expanded upon with each unsupervised observation being assigned its own class such that $\matW=\left[\begin{smallmatrix}\matC_{} & 0\\ 0 & \mathbf{I}\end{smallmatrix}\right]$. Both $\matW$ and the auxiliary matrix are then updated iteratively with that information taken into account. While this algorithm serves a different purpose than the one that will be described in Chapter~\ref{method}, it still is able to provide a contribution to the updating process described in Section~\ref{updatesec}. \cite{yang2018non} further expand upon the work of \cite{liu2011constrained} by adding in a sparsity constraint, making it a dual constrained algorithm.

\cite{arora2016computing} introduced the concept of separable nonnegative matrices, which are solved more efficiently, even in the presence of noise. A nonnegative matrix, $\matX$, is defined to be separable if there exists an index set $K$ of cardinality $r$ and a nonnegative matrix $\matH$ such that $\matX = \matX(:,K)\matH$. This concept was then expanded upon by \cite{gillis2014successive} to create the Successive Nonnegative Projection Algorithm; a fast algorithm similar to NMF that uses near-separable matrices in order to decompose the data of interest. This algorithm as well as separable nonnegative matrix factorization \citep{gillis2014robust} have been used in multiple spectral data applications \citep{qu2015subspace, luce2016using}.

\section{Methodology}\label{method}
\subsection{The Group and Basis Restricted NMF (GBR-NMF) algorithm}\label{newcon}
The conventional NMF algorithm proposed by \cite{lee1999learning} is a completely unsupervised algorithm, meaning one cannot specify any suspected factors in the model or specify any clustering constraints if there are suspected groups within the data. Here, we propose a variation of the work by \cite{liu2011constrained} by similarly utilizing a further decomposition of the data of interest into matrices $\matW$, $\matA$, and $\matS$ such that
\begin{equation}\label{maineq2}
    \matX = \matW\matA\matS + \epsilon,
\end{equation}
where it is apparent that $\matH$ from Equation~\ref{maineq} is further decomposed into a $q \times q$ auxiliary matrix $\matA$ and the $q \times p$ matrix $\matS$ containing the underlying factors. However, this algorithm differs in the way that restrictions are imposed on the matrices. \cite{liu2011constrained} restrict their $\matW$ matrix as stated in Section~\ref{prevcon}, whereas the new algorithm restricts a number of columns in $\matW$ through the use of a complete indicator matrix and also restricts some rows in $\matS$. The added restrictions allow for one to specify known groups in $\matW$ and suspected factors in $\matS$, further improving the interpretability of the  model. It is important to note that while there is further decomposition, the matrices $\matW$ and $\matS$ can still be interpreted the same way as $\matW$ and $\matH$ respectively. However, due to the fact that we will now be imposing a constraint on $\matS$, $\matA$ is needed to perform adjustments on the constrained factors. We propose that $\matA$ be initialized as an identity matrix which, through the iterative updating process, will act as a scaling matrix to the rows of $\matS$ --- a detail which will be shown to be quite significant in Section~\ref{raman}.

As previously specified, constraints will now be added to $\matS$, herein creating a restricted model. Assuming that there are $q$ underlying factors in the model, suppose there exist $k$ known and $q - k$ unknown factors in the model. Then these $k$ known factors can be specified as factors in $\matS$ that are not updated through the iterative process, while the $q - k$ factors are still included in the updating process. This allows for the algorithm to take into consideration the factors that a user may know to be present and use them to estimate other factors that may not be known to the users.

Branching from the work of \cite{ding2005equivalence}, an optional constraint regarding the $\matW$ matrix has also been added for applications regarding clustering. As previously described in Section \ref{prevcon}, basic NMF resembles a soft K-means clustering algorithm under use of the Frobenius norm as an objective function. Here, the features in $\matS$ represent the centroids in the algorithm and the scores in $\matW$ are how each observation scores with respect to these centroids. With this in mind, we have allowed the constraint of the first $g$ columns of $\matW$ to be a known grouping matrix. These columns are never updated in the model and the factors regarding these columns are kept unconstrained.

Thus, in the case where both constraints are employed we have $g+k \leq q$ with the first $g$ columns of $\matW$ and rows of $\matS$ responsible for the clustering results with the next $k$ accounting for the constrained factors. This allows for the major differences in groups to be found while other dissimilarities are found in the other $q-g$ factors.

\subsection{Updating Process}\label{updatesec}
Assuming we are updating with respect to Gaussian distributed error, we are required to update with respect to the Frobenius norm
\begin{eqnarray}\label{frobnorm}
    D_F(\matX \mid \matW\matA\matS) & = & \frac{1}{2}\mid\mid \matX-\matW\matA\matS \mid\mid^2\\ 
    & \propto & \mid\mid \matX-\matW\matA\matS \mid\mid^2
\end{eqnarray}
as the objective function of interest with the constraint that $w_{ij} \geq 0, a_{ij} \geq 0,$ and $s_{ij} \geq 0$ for all $ij \in \mathbb{R}$. As previously stated in Section \ref{nmf}, it is not feasible to find global minima as the GBR-NMF is not convex in $\matW, \matA,$ and $\matS$, which requires an updating process to instead find local minima. To ensure a suitable local minimum has been achieved, the GBR-NMF algorithm was applied to different data sets multiple times in order to ensure consistency in the accuracy of the algorithm. These different test cases included different initializations of the $\matW$, $\matA$, and $\matS$ matrices in an attempt to find different local minima. If one finds themselves in the situation where they believe the local minimum is unsatisfactory, they too could use the algorithm in multiple cases and compare Frobenius norms to find the most suitable answer. 

Applying the trace function, as well as both the cyclic and transpose properties associated with it, we have
\begin{align}
\begin{split}
    D(\matX \mid \matW\matA\matS) & \propto Tr((\matX-\matW\matA\matS)(\matX-\matW\matA\matS)^T)\\
    & = Tr(\matX\matX^T)- 2Tr(\matW\matA\matS\matX^T) + Tr(\matW\matA\matS\matS^T\matA^T\matW^T).
\end{split}
\end{align}
However, this does not account for the nonnegativity constraints previously specified. In order to do so, a Lagrange function ($\mathcal{L}$) with multipliers $\alpha_{ij}, \beta_{ij}$, and $\gamma_{ij}$ are required for constraints $w_{ij} \geq 0, a_{ij} \geq 0$ and $s_{ij} \geq 0$ respectively (note that $\boldsymbol{\alpha} = [\alpha_{ij}]$, $\boldsymbol{\beta} = [\beta_{ij}]$, and $\boldsymbol{\gamma} = [\gamma_{ij}]$). This gives the following Lagrange function
\begin{align}
\begin{split}
    \mathcal{L} = & Tr(\matX\matX^T)- 2Tr(\matW\matA\matS\matX^T) + Tr(\matW\matA\matS\matS^T\matA^T\matW^T)\\
    & + Tr(\boldsymbol{\alpha}\matW^T) + Tr(\boldsymbol{\beta}\matA^T) + Tr(\boldsymbol{\gamma}\matS^T).
\end{split}
\end{align}
Now needing the derivatives of $\mathcal{L}$ to be zero with respect to $\matW$, $\matA$, and $\matS$ we have
\begin{equation*}
    \frac{\partial \mathcal{L}}{\partial \matW} = -2\matX\matS^T\matA^T + 2\matW\matA\matS\matS^T\matA^T + \alpha = 0,
\end{equation*}
\begin{equation}
    \frac{\partial \mathcal{L}}{\partial \matA} = -2\matW^T\matX\matS^T + 2\matW^T\matW\matA\matS\matS^T + \beta = 0,
\end{equation}
\begin{equation*}
    \frac{\partial \mathcal{L}}{\partial \matS} = -2\matA^T\matW^T\matX + 2 \matA^T\matW^T\matW\matA\matS + \gamma = 0.
\end{equation*}
Then applying the Karush-Kuhn-Tucker conditions $\alpha_{ij}w_{ij} = 0$, $\beta_{ij}a_{ij} = 0$, and $\gamma_{ij}s_{ij} = 0$, we obtain
\begin{equation*}
    (\matX\matS^T\matA^T)_{ij}w_{ij} - (\matW\matA\matS\matS^T\matA^T)_{ij}w_{ij} = 0,
\end{equation*}
\begin{equation}
    (\matW^T\matX\matS^T)_{ij}a_{ij} - (\matW^T\matW\matA\matS\matS^T)_{ij}a_{ij} = 0,
\end{equation}
\begin{equation*}
    (\matA^T\matW^T\matX)_{ij}s_{ij} - (\matA^T\matW^T\matW\matA\matS)_{ij}s_{ij} = 0.
\end{equation*}
Which leads to the following updating rules
\begin{equation}\label{updateW}
    w_{ij} \leftarrow w_{ij}\frac{(\matX\matS^T\matA^T)_{ij}}{(\matW\matA\matS\matS^T\matA^T)_{ij}},
\end{equation}
\begin{equation}\label{updateA}
    a_{ij} \leftarrow a_{ij}\frac{(\matW^T\matX\matS^T)_{ij}}{(\matW^T\matW\matA\matS\matS^T)_{ij}},
\end{equation}
\begin{equation}\label{updateS}
    s_{ij} \leftarrow s_{ij}\frac{(\matA^T\matW^T\matX)_{ij}}{(\matA^T\matW^T\matW\matA\matS)_{ij}}.
\end{equation}

Further detail of the derivations in this section can be found in Appendix \ref{pua}.

\subsection{Proof of Convergence}\label{pocsection}
In order to prove that the above updates are correct, the objective function must be non-increasing according to the updates. The objective function is then invariant under the updates if and only if $\matW$, $\matA$, and $\matS$ are at stationary points \citep{liu2011constrained}. In order to prove this, we must borrow a lemma stemming from updates to the EM algorithm \citep{dempster1977maximum,wu1983convergence} provided below.
\begin{lemma}
    If an auxiliary function, $G$, exists for $F(x)$ and satisfies the conditions 
    \begin{enumerate}
        \item $G(x,x')\geq F(x)$
        \item $G(x,x) = F(x)$
    \end{enumerate}
    then $F$ is non-increasing using the update
    \begin{equation*}
        x^{t+1} = \argmin_x G(x,x')
    \end{equation*}
\end{lemma}

Given that $F_{w_{ij}}$ is defined as the parts of equation~\ref{frobnorm} relevant to $w_{ij}$ and $F_{w_{ij}}'$ is the derivative of such, we then need to prove the following three lemmas.
\begin{lemma}\label{auxW}
    The function $$G(w,w_{ij}) = F_{w_{ij}}(w_{ij}^t) + F_{w_{ij}}' (w_{ij})(w-w_{ij}^t)+\frac{(\matW\matA\matS\matS^T\matA^T)_{ij}}{w_{ij}^t}(w-w_{ij}^t)^2$$ is an auxiliary function to $F_{w_{ij}}$.
\end{lemma}

\begin{lemma}\label{auxA}
    The function $$G(a,a_{ij}) = F_{a_{ij}}(a_{ij}^t) + F_{a_{ij}}' (a_{ij})(a-a_{ij}^t)+\frac{(\matW^T\matW\matA\matS\matS^T)_{ij}}{a_{ij}^t}(a-a_{ij}^t)^2$$ is an auxiliary function to $F_{a_{ij}}$.
\end{lemma}

\begin{lemma}\label{auxS}
        The function $$G(s,s_{ij}) = F_{s_{ij}}(s_{ij}^t) + F_{s_{ij}}' (s_{ij})(s-s_{ij}^t)+\frac{(\matA^T\matW^T\matW\matA\matS)_{ij}}{s_{ij}^t}(s-s_{ij}^t)^2$$ is an auxiliary function to $F_{s_{ij}}$.
\end{lemma}

These lemmas are proved in a very similar fashion. As a result, lemma~\ref{auxW} will be proved below, while Lemmas~\ref{auxA} and \ref{auxS} can be found in appendix~\ref{poc}. With minor modifications of the proof by \cite{liu2011constrained}, the following can be proved.

\begin{proof}
 It is clear that $G(w,w) = F_{w_{ij}}(w)$. According to the definition of an auxiliary function, it only needs to be shown that $G(w,w_{ij}^t)\geq F_{w_{ij}}(w)$. This can be done using the Taylor series expansion of $F_{w_{ij}}$:
 $$F_{w_{ij}}(w) = F_{w_{ij}}(w_{ij}^t)+F_{w_{ij}}'(w-w_{ij}^t)+\frac{1}{2}F_{w_{ij}}''(w-w_{ij}^t)^2$$
 with $F_{w_{ij}}''$ being the second order derivative of $F_{w_{ij}}$. We can show that
 $$F_{w_{ij}} = (\frac{\partial D}{\partial W})_{ij} = (-2\matX\matS^T\matA^T + 2\matW\matA\matS\matS^T\matA^T)_{ij},$$
 $$F_{w_{ij}}'' = (\frac{\partial^2 D}{\partial \matW^2})_{ij} = 2(\matA\matS\matS^T\matA^T)_{jj}.$$
 As $F_{w_{ij}}''$ has now been solved for, it is equivalent to prove
 \begin{equation}\label{midproof}
 \frac{(\matW\matA\matS\matS^T\matA^T)_{ij}}{w_{ij}^t} \geq \frac{1}{2}F_{ij}'' = (\matA\matS\matS^T\matA^T)_{jj}
 \end{equation}
in order to reach Lemma~\ref{auxW}. Now we have
 \begin{eqnarray}
     (\matW\matA\matS\matS^T\matA^T)_{ij} & = & \sum_{\ell=1}^q (\matW)_{i\ell}(\matA\matS\matS^T\matA^T)_{\ell j}\\
     & \geq & w_{ij}^t(\matA\matS\matS^T\matA)_{jj}
 \end{eqnarray}
which when rearranged proves Equation~\ref{midproof}.
\end{proof}

\subsection{Algorithm Summary}
For the observed $n \times p$ data matrix, $\matX$, with $c$ known groups in $\matW$, $k$ suspected factors in $\matS$ and $q$ total factors, the algorithm can be summarized using the following steps:
\begin{enumerate}
    \item If an $n \times c$ cluster membership matrix is specified, set the first $c$ columns of the matrix, $\matW$, to be this matrix. Initialize the remaining $n \times (q-c)$ elements sampling from the uniform distribution $U(\min(\matX),\max(\matX))$. Otherwise, initialize the entire matrix by sampling from the uniform distribution specified above.
    \item Initialize $\matA$ such that it takes the form of a $q \times q$ identity matrix.
    \item Initialize the remaining $(q-k-c)$ columns of $\matS$ sampling from the uniform distribution on the bounds of the known factor values.
    \item Calculate $D_F^{(0)}$ using equation~\ref{frobnorm}.
    \item \label{loop} Set $z=1$ and enter the updating process:
    \begin{enumerate}
        \item \label{firstloop}Update the unspecified elements of $\matW$ using Equation~\ref{updateW}.
        \item Update all elements of $\matA$ using Equation~\ref{updateA}.
        \item Update the unspecified elements of $\matS$ using Equation~\ref{updateS}.
        \item Calculate $D_F^{(i)}$ using equation~\ref{frobnorm}.
        \item \label{convcheck} If $D_F^{(z-1)}-D_F^{(z)}$ is less than a specified $\delta$ value or the value of $i$ is equal to the maximum iteration count, move to step~\ref{return}. Otherwise set $z = z+1$ and return to step~\ref{firstloop}.
    \end{enumerate}
    \item \label{return} Return the updated values of $\matW$, $\matA$, and $\matS$.
\end{enumerate}

It is important to note that stopping criterion of step~\ref{convcheck} can be determined in a variety of ways. GBR-NMF provides two possible methods which include lack of progress and a maximum iteration count. Unless otherwise specified, the maximum iteration count is used as lack of progress depends greatly on the value of $\delta$ provided. Data sets with large variability may not be able to attain a small epsilon value. Contrarily, if the specified $\delta$ value is too large then the stopping criterion will be achieved too quickly, leaving room for improvement. To battle this, a high maximum iteration count of 50,000 is set to ensure a local maximum is reached. 

\section{Applications}

\subsection{Simulated Data}
A simulation was generated in order to test the performance of the algorithm using $n=400$ observations, $p=2000$ variables, $g=4$ groups, and $q=7$ true underlying factors. This simulation was created to test the accuracy with respect to the clustering aspect of the algorithm as well as how it does predicting scores on known factors.
%
%
The simulation stipulates that each group differentiates itself from the others by having one unique factor that the others did not. Each one of the four groups had a unique factor, along with three other factors that were common with the other groups.  Thus, when running the algorithm with respect to this data, the first four columns of the $\matW$ matrix are set to be the known classification matrix of the model. To add to the simulations, one of the three common features in $\matS$ were held as a constrained factor to determine how well it's true scores were being found. The matrices below further illustrate which columns and rows in the model are being held constant (in bold) throughout the updating process.

\begin{equation}
\matW=
    \begin{pmatrix}
        \mathbf{1} & \mathbf{0} & \mathbf{0} & \mathbf{0} & w_{1,5} & w_{1,6} & w_{1,7} \\
        \mathbf{0} & \mathbf{1} & \mathbf{0} & \mathbf{0} & w_{2,5} & w_{2,6} & w_{2,7} \\
        \mathbf{0} & \mathbf{0} & \mathbf{1} & \mathbf{0} & w_{3,5} & w_{3,6} & w_{3,7} \\
        \mathbf{0} & \mathbf{0} & \mathbf{0} & \mathbf{1} & w_{4,5} & w_{4,6} & w_{4,7}. \\
        \mathbf{1} & \mathbf{0} & \mathbf{0} & \mathbf{0} & w_{5,5} & w_{5,6} & w_{5,7} \\
        \mathbf{\vdots} & \mathbf{\vdots} & \mathbf{\vdots} & \mathbf{\vdots} & \vdots & \vdots & \vdots \\
        \mathbf{0} & \mathbf{0} & \mathbf{0} & \mathbf{1} & w_{400,5} & w_{400,6} & w_{400,7} \\
    \end{pmatrix}
\end{equation}
\begin{equation}
\matA=
    \begin{pmatrix}
        a_{1,1} & 0 & 0 & 0 & 0 & 0 & 0 \\
        0 & a_{2,2} & 0 & 0 & 0 & 0 & 0 \\
        0 & 0 & a_{3,3} & 0 & 0 & 0 & 0 \\
        0 & 0 & 0 & a_{4,4} & 0 & 0 & 0 \\
        0 & 0 & 0 & 0 & a_{5,5} & 0 & 0 \\
        0 & 0 & 0 & 0 & 0 & a_{6,6} & 0 \\
        0 & 0 & 0 & 0 & 0 & 0 & a_{7,7} \\
    \end{pmatrix}
\end{equation}
\begin{equation}
\matS=
    \begin{pmatrix}
        s_{1,1} & s_{1,2} & \ldots & s_{1,2000} \\
        s_{2,1} & s_{2,2} & \ldots & s_{2,2000} \\
        s_{3,1} & s_{3,2} & \ldots & s_{3,2000} \\
        s_{4,1} & s_{4,2} & \ldots & s_{4,2000} \\
        \mathbf{s_{5,1}} & \mathbf{s_{5,2}} & \mathbf{\ldots} & \mathbf{s_{5,2000}} \\
        s_{6,1} & s_{6,2} & \ldots & s_{6,2000} \\
        s_{7,1} & s_{7,2} & \ldots & s_{7,2000} \\
    \end{pmatrix}
\end{equation}
Entries for $\matW$, $\matA$, and $\matS$ were all created using random uniform distributions and then scaled accordingly such that the column sums of $\matW$ are equal to $n$ and the area under the curve of the columns of $\matS$ were equal to 1. Thus the diagonal elements of $\matA$ were scaled inversely such that the overall reconstructions would remain the same as those prior to scaling.

These data sets were created and decomposed via the GBR-NMF 100 different times in order to ensure accuracy of the results. The residual sum of squares (RSS) comparing the true $\matS$ with the estimated $\matS$ were then calculated and a mean and standard error of these numbers were taken across all features (found in Table~\ref{tab:simdata1}). This was similarly done with the resulting scores matrix $\matW$ and auxiliary matrix $\matA$ to quantify the algorithm's overall estimation performance. More specifically, to determine how close the values of the calculated matrices are to the true values. Table~\ref{tab:simdata1} displays that the algorithm does exceptionally well in predicting the features in both simulations. Meanwhile, the RSS of the scores matrix was also smaller in comparison to the RSS regarding the $\matA$ matrix. This is primarily due to the scalings on the two outside matrices which causes the values to become quite close. Therefore, a more fitting comparison would be to the unconstrained model, which is also displayed in Table~\ref{tab:simdata1}. This shows a drastic increase in both the accuracy of the scores and factors in comparison to the original model. It is important to note that this is a direct comparison between the standard NMF model and GBR-NMF. Comparisons to PCA or the semi-supervised NMF model of \cite{liu2011constrained} would be inaccurate as a direct comparison is not available. Principal components acquired from PCA are uncorrelated with respect to one another, making it difficult to make a direct comparison to the true factors created. Similarly, the different constraints in the work of \cite{liu2011constrained} create different interpretations of the factor matrices.

\begin{table}
    \centering
    \begin{tabular}{| c | c |  c |}
        \hline
        \multirow{2}{*}{$\matW$ constraint held} & \multicolumn{2}{c|}{Standard NMF}\\
         & Average & Standard Error \\
        \hline
        $\matW$ RSS & \ 4.7 e\ \  +9 & 1.9 e +10 \\
        $\matH$ RSS & 4.0 e\ \ -5 & 1.0 e\ \ -5\ \ \ \\
        \hline
        \multirow{2}{*}{$\matW$ constraint held} & \multicolumn{2}{c|}{GBR-NMF}\\
         & Average & Standard Error \\
        \hline
        $\matW\matA$ RSS &  \ 2.6 e\ \ +9 & \ 7.3 e\ \ +8\ \ \ \\
        $\matS$ RSS & 8.0 e\ \ -6 & 4.0 e\ \ -6\ \ \  \\
        \hline
    \end{tabular}
    \caption{Summary of the residual sum of squares (RSS) of both the learned features and scores versus the true features and scores for both standard NMF and GBR-NMF.}
    \label{tab:simdata1}
\end{table}

\subsection{Facial Expression Data}
\begin{figure}
    \center{
    \begin{tabular}{c}
         \includegraphics[width=4.8in]{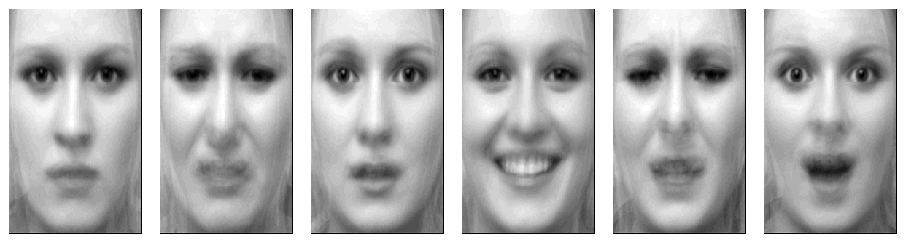} \\
         \includegraphics[width=4.8in]{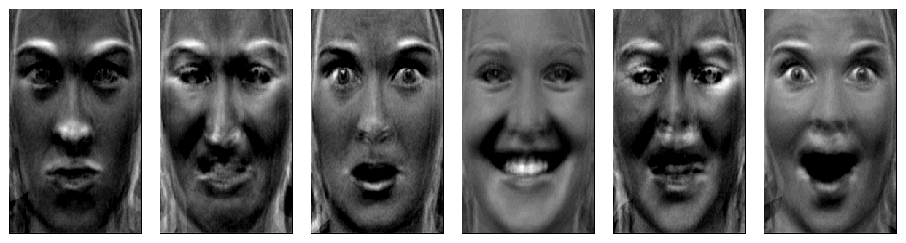}\\
         \includegraphics[width=4.8in]{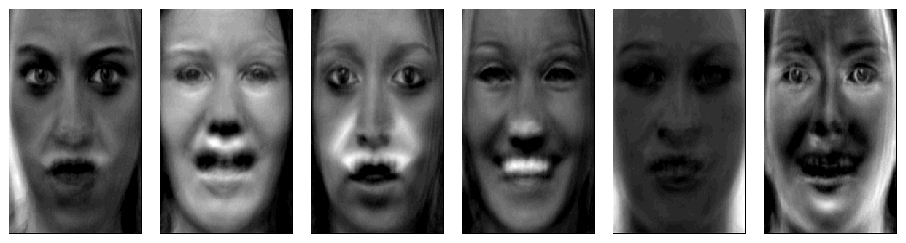}
    \end{tabular}
        
    }
    \caption{Row one: Six facial expressions (neutral excluded) averaged over the twelve different women. The facial expressions, in order, are anger, disgust, fear, happiness, pain, and surprise. Row two: Six retrieved unconstrained facial features with respect to grouping matrix corresponding to expressions from the GBR-NMF. Row three: Six retrieved features from the completely unconstrained basic NMF model.}
    \label{fig:avgfaces}
\end{figure}

A common research application of nonnegative matrix factorization is data regarding facial expressions \citep{buciu2004application,guillamet2002classifying,zhi2010graph} typically for the identification of certain facial expressions or people. The pain expression data set \citep{pics2019} consists of 84 observations of $241 \times 181$ pixel facial images. These observations can be further subsetted into the faces of 12 different women; each of which displaying 7 different facial expressions. An average face of each facial expression (aside from neutral) is displayed in row one of Figure~\ref{fig:avgfaces}.
These six facial expressions are the grouping variable of interest in this data set and are placed as a grouping constraint in the $\matW$ matrix in the GBR-NMF. Along with the grouping constraints, twelve features were also constrained to be the neutral facial expressions of the different women. These constraints are determined to test what bases are acquired from the grouping variables as well as how close one can get to the true facial expression using a person's neutral face and grouping information. From these neutral faces, the other bases are recovered and built on top of this neutral face in order to recover the original images. The six bases recovered using GBR-NMF resemble key facial features explanatory of each emotion. Row two of Figure~\ref{fig:avgfaces} identifies the unconstrained features retrieved from the model. The fourth image of the second row displays a smile expression, which is portrayed well by highlighting the teeth and perked cheeks. These results are more interpretable and separated than those that are recovered from the standard NMF algorithm as this algorithm tends to recover features that appear to more closely resemble particular women's faces rather than emotion-specific representations. The standard NMF results are shown in row three of Figure~\ref{fig:avgfaces} which displays the six features of an unconstrained NMF sorted to best align with the original expressions. 



\begin{figure}
    \centering
    \includegraphics[width =3in]{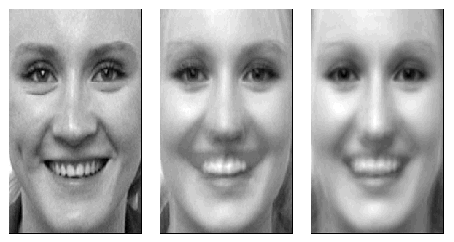}
    \caption{An original picture of a woman's face (left) along with reconstructions from the GBR-NMF algorithm (center) and the standard NMF algorithm (right).}
    \label{fig:face_recons}
\end{figure}

This application displays that we are acquiring better separation in the GBR-NMF in comparison to the basic NMF with respect to the recovered facial expressions. These interpretability benefits could be assumed to result in a decrease in overall model performance. However, as demonstrated in Figure~\ref{fig:face_recons}, the reconstructions still do compare quite well to those of the unconstrained NMF model, making this a worthy trade-off.

\subsubsection{Predicting an Unseen Facial Expression}
\begin{figure}
    \centering
    \includegraphics[width=3in]{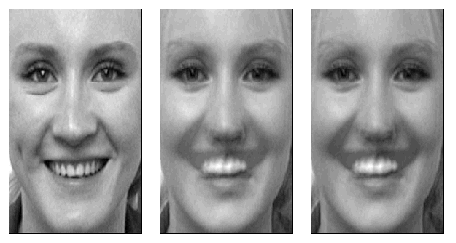}
    \caption{True facial expression (left) in comparison to GBR-NMF prediction of single missing facial expression (center) and GBR-NMF missing facial expression prediction using only neutral face (right).}
    \label{fig:face_predictions}
\end{figure}

In this section, we consider the possibility of generating previously unseen images using the GBR-NMF algorithm. Suppose one wanted to predict a person's facial expression that is not included in the original model fit. Then the GBR-NMF could be used in order to predict what that expression might look like. The only information that would be needed in order to do so are the acquired bases from the model, as well as one's neutral facial expression (ie. the $\matS$ matrix). This problem can be tackled in two different ways. The first scenario involves simply removing one observation in the data set and attempting to predict the face that was removed. The second requires all of the data regarding a person's facial expressions to be removed from the model and then predicting what their expression would look like; again, using only the neutral facial expression and the bases acquired. The predictions of these faces were made by scoring 1 on the desired facial expression and 1 on the neutral face, and using this linear combination as the prediction. Acquisitions of the predicted faces, along with the true facial expression are shown in Figure~\ref{fig:face_predictions}. While it is clear that these reconstructions are not as close to the original image as the results in Figure~\ref{fig:face_recons}, they illustrate a predictive capability for the semi-supervised approach that is not achievable through the conventional unsupervised approach.

\subsection{Raman Spectroscopic Cancer Data}\label{raman}
\begin{figure}
    \center{
        \includegraphics[width=4.8in]{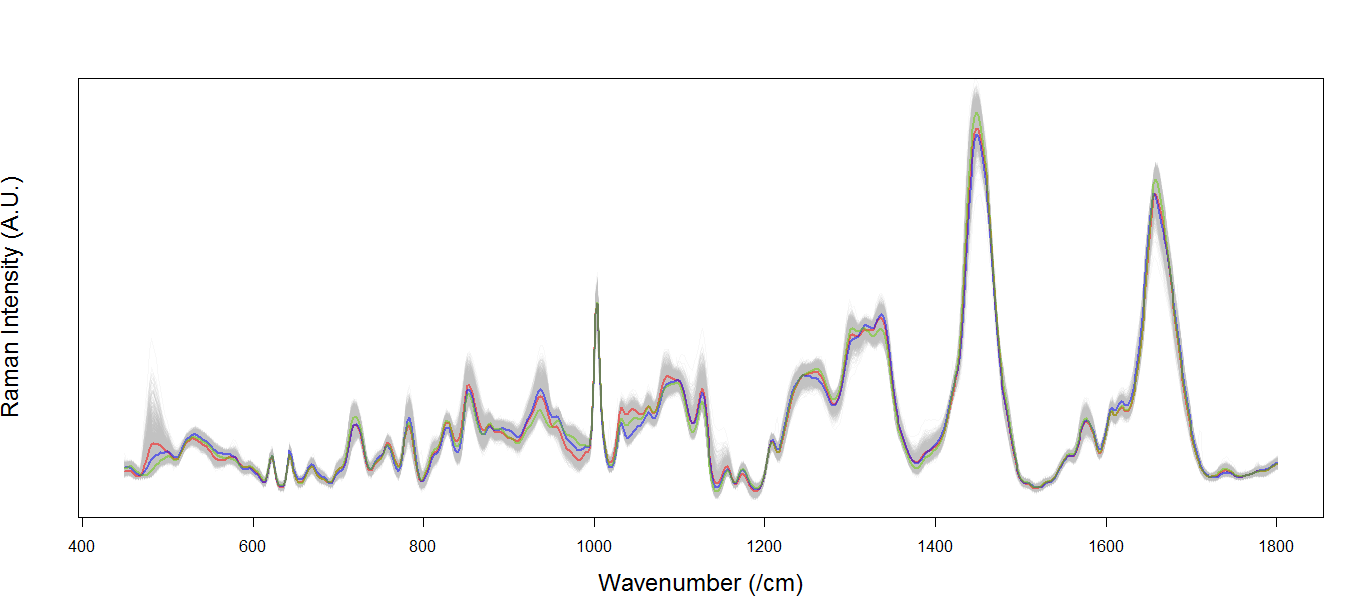}
    }
    \caption{Three averaged Raman spectra displaying three different cell lines which include H460 (red), MCF-7 (blue), LNCaP (green) tumor cells. Grey lines are the observed Raman spectra. Unit displayed on the y-axis is arbitrary intensity (AI).}
    \label{fig:cell_lines}
\end{figure}

In this example, Raman spectroscopy can be used to provide detailed information on the changes in relative concentrations of classes of molecular compounds (proteins, lipids, DNA, metabolites, etc.) within cellular and tissue environments exposed to ionizing radiation used in cancer therapy \citep{matthews2015radiation,harder2015raman,matthews2010variability,paidi2019label}. \cite{matthews2015radiation} introduce data containing 3240 spectra containing 582 Raman intensities (arbitrary units) at different wavenumbers ($cm^{-1}$). These spectra are separated into three different groups based on cell type, specifically lung (H460), breast (MCF-7), and prostate (LNCaP) tumours. The spectra, overlaid with an average of each cell type, can be seen in the Figure~\ref{fig:cell_lines}.
In Raman spectroscopy, chemical constituents are identified through comparison with known chemical constituent spectra and literature line lists exist for the most common cellular components.
\cite{matthews2015radiation} performed a principal component analysis on these spectra and found that the cellular response to radiation exhibited a strong glycogen signal, identified as the 1st principal component in the data. The aim of this study is to further expand on the work of \cite{matthews2015radiation} by recovering more biological bases using GBR-NMF. Using the GBR-NMF allows for the constraint of the factor matrix, ensuring the previously recovered glycogen is specified in the model. Further recovered bases are then analyzed to discover new biological information with respect to the chemical composition of the cells in the data set.

There are two main, and related, disadvantages to using PCA in this context. Firstly, permitting negative scores and component coefficients violates the physical understanding of the chemical composition of a cell --- reconstructions of the spectra could be built in a subtractive manner where an overabundance of one chemical is estimated and then subtracted via a negative score on another chemical. Secondly, and perhaps related to the problems with previous assumptions, \cite{matthews2015radiation} found a lack interpretability in additional principal components of the model. The standard NMF algorithm gains a key advantage over PCA due to the fact that it is forced to be an additive model with respect to chemical constituents, which matches the physical construction of Raman spectroscopic data. This is demonstrated by \cite{luce2016using} using the Succesive Nonnegative Projection Algorithm to decompose time-resolved Raman spectroscopy into constituent spectra and kinetics of the underlying reactions. However, this method is only presented as an unsupervised learning technique and restricts one from being able to specify previously suspected constituent spectra. \cite{deng2020monitor} further illustrates this point by performing unconstrained NMF on the same data set and not only discovering glycogen, but also a spectrum very similar to that of a lipid.  While using standard NMF is beneficial over PCA, use of the GBR-NMF allows making use of previous results from \cite{matthews2015radiation} wherein an important chemical, glycogen, was discovered and biologically confirmed. It also ensures that the data will be decomposed into a linear combination of predominantly known features, which is something that cannot be guaranteed by conventional. As such, GBR-NMF is utilized by constraining an acquired spectra of pure glycogen within $\matS$, and further constrain $\matW$ with the known cell lines in the data. We permit one further basis row and group column to be fully unconstrained.

\begin{figure}
    \centering
    \includegraphics[width=4.8in]{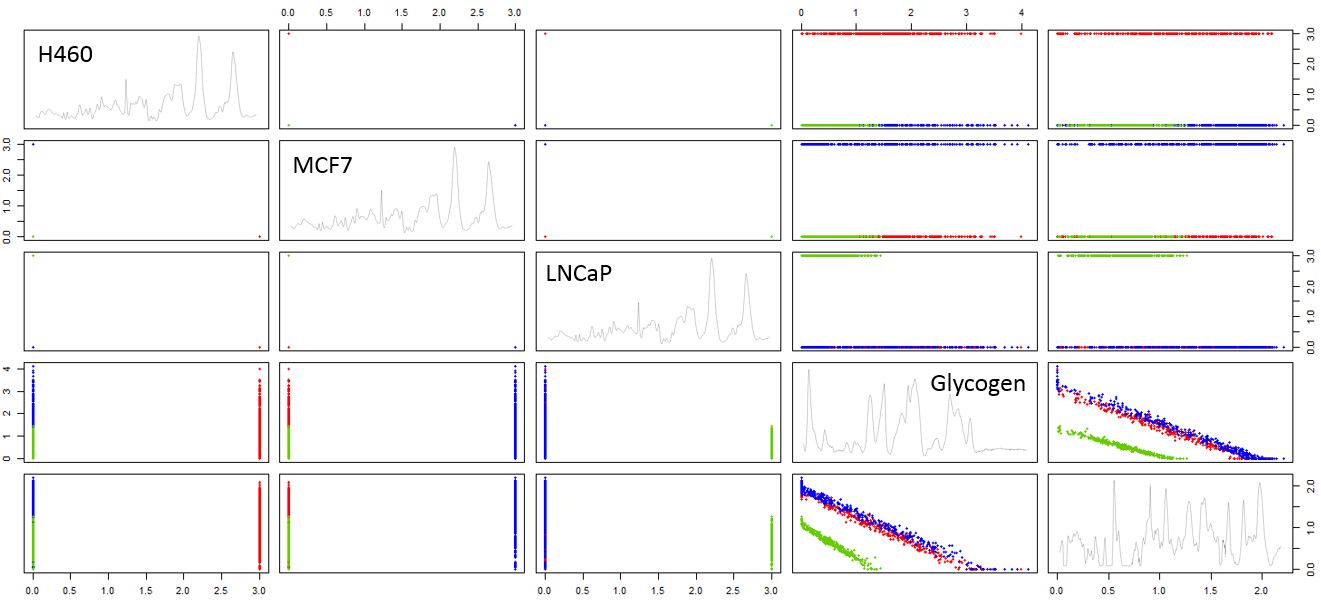}
    \caption{Scores on the five (four known/one unknown) chemical bases with respect to the three different cell lines (lung (red), breast (blue), prostate (green)). The Raman spectra in the bottom right corner is that of the unknown feature.}
    \label{fig:scatterplot}
\end{figure}

Figure~\ref{fig:scatterplot} provides a scatterplot matrix of the scores from $\matW$ on the off diagonal and the recovered spectral bases along the diagonal. 
The first three bases along the diagonal were recovered under group constraints on the three tumour types. We note that all three bases in these cases closely approximate a standard cell spectra, with slight variations among some wavelengths. The glycogen scores for the green group (prostate, LNCaP) largely fall below a score of 1.5, which cannot be said about the other two cell lines. This further supports the result put forward by \cite{matthews2015radiation} that LNCaP prostate tumours exhibit lower glycogen production in response to radiation than the other two cell lines (H460, MCF7). Furthermore, LNCaP prostate tumours are also known to be more sensitive to ionizing radiation than those of H460 and MCF-7. Interestingly, the estimated results from GBR-NMF show that the prostate cells are not fully differentiable from the other cell lines based on glycogen alone, as there is substantial overlap in the glycogen scores, even if the LNCaP cells have lower, on average, scores. However, if we consider the fifth, fully unconstrained basis that was recovered (fifth diagonal along Figure~\ref{fig:scatterplot}), we also see lower scores on average from the LNCaP cells. In fact, if the scores are viewed in tandem between glycogen and this currently unknown spectral component, we see complete separation between LNCaP cells (green) and the H460 and MCF-7 cell lines (blue and red). Thus, applying the GBR-NMF method has provided a new spectral component to investigate with regards to potential chemicals that could be indicative of cell radiosensitivity. The current result presented here allows for a more direct focus on individual biochemical cellular response to radiotherapy as compared with the original basic PCA analysis. Further work is required to elucidate the biochemistry of radiation response and is beyond the scope of this manuscript.

\begin{figure}
    \center{
        \includegraphics[width=4.8in]{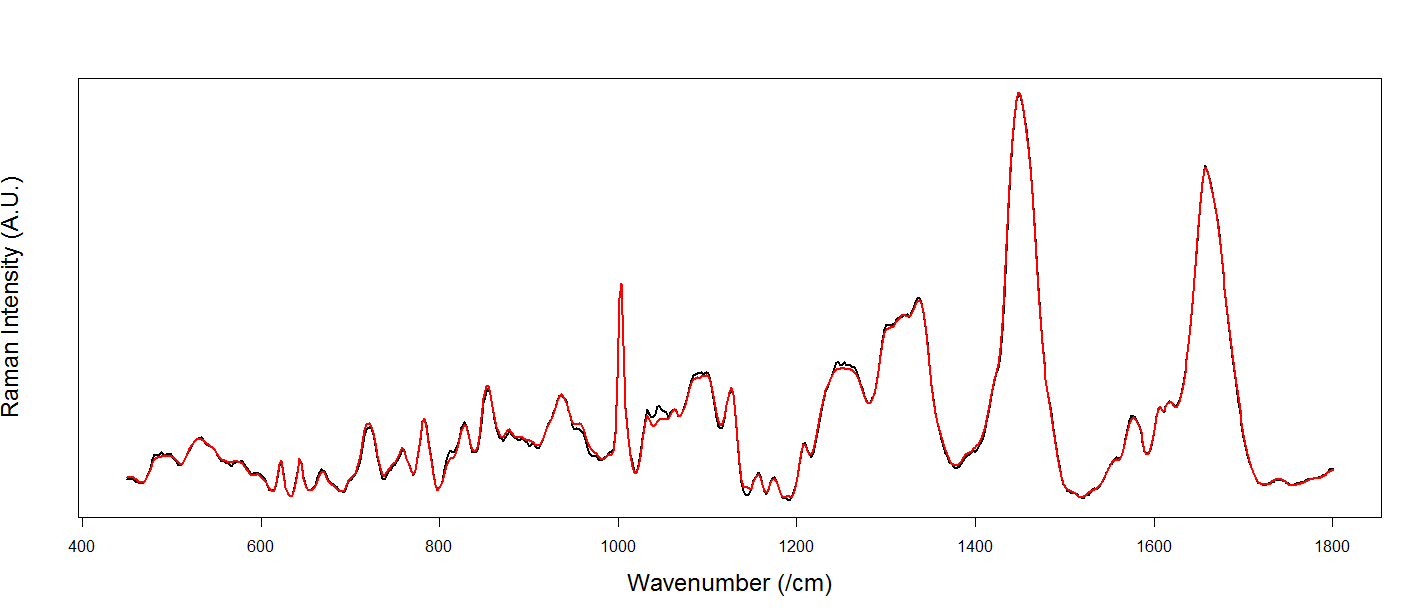}
    }
    \caption{An experimentally derived Raman spectrum (black) versus a reconstructed spectrum (red) from the analysis summarized graphically in Figure~\ref{fig:scatterplot}.}
    \label{fig:reconstruction}
\end{figure}

A natural question that arises is how well the model estimated by GBR-NMF approximates the original spectra. Consider the plot in Figure~\ref{fig:reconstruction} showing two spectra, one being a randomly selected spectrum from the original data and the other being a reconstruction of that spectrum from GBR-NMF. From a spectroscopic viewpoint, this would be considered a suitable reconstruction.

\section{Summary}
A novel group and basis restricted nonnegative matrix factorization (GBR-NMF) algorithm  was developed to input known bases and/or known groups within a nonnegative matrix factorization model. Updates were formulated and then tested via simulations and applications to real data: specifically, digital facial images and Raman spectroscopy. GBR-NMF was used to specify known to exist a priori factors and groups in the models. Regarding the facial expression data, GBR-NMF displayed significant improvements on standard NMF with respect to the interpretability of the factors discovered under the group constraints. For the spectral data, the semi-supervised approach permitted by the GBR-NMF framework can aide the scientific process by allowing pre-specification of known constituent chemicals and simultaneously suggesting potential constituents through the unconstrained factors.

Future methodological work could consider adjustments made to the auxiliary matrix such as adjusting the off-diagonal elements to be of non-zero value. Further exploration of $\epsilon$ in Equation~\ref{maineq2} should also be considered. Namely, the development of new updating procedures in regards to different noise distributions and how the algorithm is affected when the assumption of Gaussian distributed error is violated. On the application side, the authors and collaborators are currently investigating the unknown Raman spectrum acquired (fifth diagonal of Figure~\ref{fig:scatterplot}) from applying GBR-NMF to the data from \cite{matthews2015radiation} for potential biological importance in indicating tumour radiosensitivity --- such work has important ramifications for personalized radiotherapy programs. Research regarding error distributions of Raman spectroscopic data sets should also be investigated further, as the process of baseline subtraction and setting area under the curve of spectra to equal 1 could drastically affect this term.

\newpage
\bibliographystyle{chicago}  
\bibliography{references}

\begin{appendices}
\section{Derivation of Updating Algorithm}\label{pua}
\begin{eqnarray*}
    D_F & = & \frac{1}{2}\mid\mid \matX - \matW\matA\matS \mid\mid^2 \\
    & \propto & \mid\mid \matX - \matW\matA\matS \mid\mid^2\\
    & = & (\sqrt{Tr((\matX - \matW\matA\matS)(\matX - \matW\matA\matS)^T)})^2\\
    & = & Tr((\matX - \matW\matA\matS)(\matX - \matW\matA\matS)^T)\\
    & = & Tr((\matX - \matW\matA\matS)(\matX^T - \matS^T\matA^T\matW^T))\\
    & = & Tr(\matX\matX^T-\matX\matS^T\matA^T\matW^T-\matW\matA\matS\matX^T+\matW\matA\matS\matS^T\matA^T\matW^T)\\
    & = & Tr(\matX\matX^T)-Tr(\matX\matS^T\matA^T\matW^T)-Tr(\matW\matA\matS\matX^T)\\
    & & +Tr(\matW\matA\matS\matS^T\matA^T\matW^T)\\
    & = & Tr(\matX\matX^T)-Tr((\matX\matS^T\matA^T\matW^T)^T)-Tr(\matW\matA\matS\matX^T)\\
    & & +Tr(\matW\matA\matS\matS^T\matA^T\matW^T)\\
    & = & Tr(\matX\matX^T)-Tr(\matW\matA\matS\matX^T)-Tr(\matW\matA\matS\matX^T)\\
    & & +Tr(\matW\matA\matS\matS^T\matA^T\matW^T)\\
    & = &  Tr(\matX\matX^T)-2Tr(\matW\matA\matS\matX^T)+Tr(\matW\matA\matS\matS^T\matA^T\matW^T)
\end{eqnarray*}

As specified in section \ref{updatesec}, this does not account for the nonnegativity constraints. To do so, a Lagrange function $\mathcal{L}$ with multipliers $\alpha_{ij}$, $\beta_{ij}$, and $\gamma_{ij}$ are required in order to constrain $\matW$, $\matA$, and $\matS$ such that $w_{ij} \geq 0, a_{ij} \geq 0$ and $s_{ij} \geq 0$ respectively. This gives the following Lagrange function (with $\boldsymbol{\alpha} = [\alpha_{ij}]$, $\boldsymbol{\beta} = [\beta_{ij}]$, and $\boldsymbol{\gamma} = [\gamma_{ij}]$)

\begin{eqnarray*}
    \mathcal{L} & = & Tr(\matX\matX^T)- 2Tr(\matW\matA\matS\matX^T) + Tr(\matW\matA\matS\matS^T\matA^T\matW^T)\\
    & & + Tr(\boldsymbol{\alpha}\matW^T) + Tr(\boldsymbol{\beta}\matA^T) + Tr(\boldsymbol{\gamma}\matS^T)
\end{eqnarray*}

Now finding the derivatives of $\mathcal{L}$ with respect to $\matW$, $\matA$, and $\matS$ we have

\begin{eqnarray*}
        \frac{\partial \mathcal{L}}{\partial \matW} & = & \frac{\partial}{\partial \matW}(Tr(\matX\matX^T)- 2Tr(\matW\matA\matS\matX^T) + Tr(\matW\matA\matS\matS^T\matA^T\matW^T)\\
        & & + Tr(\boldsymbol{\alpha}\matW^T) + Tr(\boldsymbol{\beta}\matA^T) + Tr(\boldsymbol{\gamma}\matS^T)) \\
        & = & \frac{\partial}{\partial \matW}Tr(\matX\matX^T)- 2\frac{\partial}{\partial \matW}Tr(\matW\matA\matS\matX^T) + \frac{\partial}{\partial \matW}Tr(\matW\matA\matS\matS^T\matA^T\matW^T)\\
        & & + \frac{\partial}{\partial \matW}Tr(\boldsymbol{\alpha}\matW^T) + \frac{\partial}{\partial \matW}Tr(\boldsymbol{\beta}\matA^T) + \frac{\partial}{\partial \matW}Tr(\boldsymbol{\gamma}\matS^T)\\
        & = & - 2\frac{\partial}{\partial \matW}Tr(\matW\matA\matS\matX^T) + \frac{\partial}{\partial \matW}Tr(\matW\matA\matS\matS^T\matA^T\matW^T) + \frac{\partial}{\partial \matW}Tr(\boldsymbol{\alpha}\matW^T)\\
        & = & -2\frac{\partial}{\partial \matW}Tr(\matX\matS^T\matA^T\matW^T) + \frac{\partial}{\partial \matW}Tr(\matW\matA\matS\matS^T\matA^T\matW^T) + \frac{\partial}{\partial \matW}Tr(\boldsymbol{\alpha}\matW^T)\\
        & = & -2\matX\matS^T\matA^T + 2\matW\matA\matS\matS^T\matA^T + \alpha\\
        & = & 0
\end{eqnarray*}

\begin{eqnarray*}
    \frac{\partial \mathcal{L}}{\partial \matA} & = & \frac{\partial}{\partial \matA}(Tr(\matX\matX^T)- 2Tr(\matW\matA\matS\matX^T) + Tr(\matW\matA\matS\matS^T\matA^T\matW^T)\\
    & & + Tr(\boldsymbol{\alpha}\matW^T) + Tr(\boldsymbol{\beta}\matA^T) + Tr(\boldsymbol{\gamma}\matS^T))\\
    & = & \frac{\partial}{\partial \matA}Tr(\matX\matX^T)- 2\frac{\partial}{\partial \matA}Tr(\matW\matA\matS\matX^T) + \frac{\partial}{\partial \matA}Tr(\matW\matA\matS\matS^T\matA^T\matW^T)\\
    & & + \frac{\partial}{\partial \matA}Tr(\boldsymbol{\alpha}\matW^T) + \frac{\partial}{\partial \matA}Tr(\boldsymbol{\beta}\matA^T) + \frac{\partial}{\partial \matA}Tr(\boldsymbol{\gamma}\matS^T)\\
    & = & - 2\frac{\partial}{\partial \matA}Tr(\matW\matA\matS\matX^T) + \frac{\partial}{\partial \matA}Tr(\matW\matA\matS\matS^T\matA^T\matW^T)\frac{\partial}{\partial \matA}Tr(\boldsymbol{\beta}\matA^T)\\
    & = & - 2\frac{\partial}{\partial \matA}Tr(\matW^T\matX\matS^T\matA^T) + \frac{\partial}{\partial \matA}Tr(\matW^T\matW\matA\matS\matS^T\matA^T)\frac{\partial}{\partial \matA}Tr(\boldsymbol{\beta}\matA^T)\\
    & = & -2\matW^T\matX\matS^T + 2\matW^T\matW\matA\matS\matS^T + \beta\\
    & = & 0
\end{eqnarray*}

\begin{eqnarray*}
    \frac{\partial \mathcal{L}}{\partial \matS} & = & \frac{\partial}{\partial \matS}(Tr(\matX\matX^T)- 2Tr(\matW\matA\matS\matX^T) + Tr(\matW\matA\matS\matS^T\matA^T\matW^T)\\
    & & + Tr(\boldsymbol{\alpha}\matW^T) + Tr(\boldsymbol{\beta}\matA^T) + Tr(\boldsymbol{\gamma}\matS^T))\\
    & = & \frac{\partial}{\partial \matS}Tr(\matX\matX^T)- 2\frac{\partial}{\partial \matS}Tr(\matW\matA\matS\matX^T) + \frac{\partial}{\partial \matS}Tr(\matW\matA\matS\matS^T\matA^T\matW^T)\\
    & & + \frac{\partial}{\partial \matS}Tr(\boldsymbol{\alpha}\matW^T) + \frac{\partial}{\partial \matS}Tr(\boldsymbol{\beta}\matA^T) + \frac{\partial}{\partial \matS}Tr(\boldsymbol{\gamma}\matS^T)\\
    & = & - 2\frac{\partial}{\partial \matS}Tr(\matW\matA\matS\matX^T) + \frac{\partial}{\partial \matS}Tr(\matW\matA\matS\matS^T\matA^T\matW^T) + \frac{\partial}{\partial \matS}Tr(\boldsymbol{\gamma}\matS^T)\\
    & = & - 2\frac{\partial}{\partial \matS}Tr(\matA^T\matW^T\matX\matS^T) + \frac{\partial}{\partial \matS}Tr(\matA^T\matW^T\matW\matA\matS\matS^T) + \frac{\partial}{\partial \matS}Tr(\boldsymbol{\gamma}\matS^T)\\
    & = & -2\matA^T\matW^T\matX + 2 \matA^T\matW^T\matW\matA\matS + \gamma\\
    & = & 0
\end{eqnarray*}

Each of these are set to equal zero as the goal is to find a local optimum of $D_F$ since it is unfeasible to find the global minimum. Applying the Karush-Kuhn-Tucker conditions,  $\alpha_{ij}w_{ij} = 0$, $\beta_{ij}a_{ij} = 0$, and $\gamma_{ij}s_{ij} = 0$ to the three equations results in the following

\begin{equation*}
    (\matX\matS^T\matA^T)_{ij}w_{ij} - (\matW\matA\matS\matS^T\matA^T)_{ij}w_{ij} = 0
\end{equation*}
\begin{equation}
    (\matW^T\matX\matS^T)_{ij}a_{ij} - (\matW^T\matW\matA\matS\matS^T)_{ij}a_{ij} = 0
\end{equation}
\begin{equation*}
    (\matA^T\matW^T\matX)_{ij}s_{ij} - (\matA^T\matW^T\matW\matA\matS)_{ij}s_{ij} = 0
\end{equation*}

which leads to the following updating rules

\begin{equation}\label{updates}
\begin{split}
    w_{ij} \leftarrow w_{ij}\frac{(\matX\matS^T\matA^T)_{ij}}{(\matW\matA\matS\matS^T\matA^T)_{ij}}\\
    a_{ij} \leftarrow a_{ij}\frac{(\matW^T\matX\matS^T)_{ij}}{(\matW^T\matW\matA\matS\matS^T)_{ij}}\\
    s_{ij} \leftarrow s_{ij}\frac{(\matA^T\matW^T\matX)_{ij}}{(\matA^T\matW^T\matW\matA\matS)_{ij}}
\end{split}
\end{equation}

\newpage

\section{Proof of Algorithm Convergence}\label{poc}
Appendix~\ref{poc} proves Lemmas~\ref{auxA} and \ref{auxS} in order to fulfill the proof requirements of Section~\ref{pocsection}.
\subsection{Proof of Lemma~\ref{auxA}}
\begin{proof}
 It is clear that $G(a,a) = F_{a_{ij}}(a)$. According to the definition of an auxiliary function, it only needs to be shown that $G(a,a_{ij}^t)\geq F_{a_{ij}}(a)$. This can be done using the Taylor series expansion of $F_{a_{ij}}$:
 $$F_{a_{ij}}(a) = F_{a_{ij}}(a_{ij}^t)+F_{a_{ij}}'(a-a_{ij}^t)+\frac{1}{2}F_{a_{ij}}''(a-a_{ij}^t)^2$$
 with $F_{a_{ij}}''$ being the second order derivative of $F_{a_{ij}}$. We can show that
 $$F_{a_{ij}} = (\frac{\partial D}{\partial A})_{ij} = (-2\matW^T\matX\matS^T + 2\matW^T\matW\matA\matS\matS^T)_{ij}$$
 $$F_{a_{ij}}'' = (\frac{\partial^2 D}{\partial \matA^2})_{ij} = 2(\matW^T\matW)_{ii}(\matS\matS^T)_{jj}$$
 It is now equivalent to prove
 \begin{equation}\label{midproof2}
 \frac{(\matW^T\matW\matA\matS\matS^T)_{ij}}{a_{ij}^t} \geq \frac{1}{2}F_{ij}'' = (\matW^T\matW)_{ii}(\matS\matS^T)_{jj}
 \end{equation}
 Now we have
 \begin{eqnarray}
     (\matW^T\matW\matA\matS\matS^T)_{ij} & = & \sum_{\ell=1}^q (\matW^T\matW\matA)_{i\ell}(\matS\matS^T)_{\ell j}\\
     & \geq & (\matW^T\matW\matA)_{ij}(\matS\matS^T)_{jj}\\
     & \geq & \sum_{\ell=1}^q(\matW^T\matW)_{i\ell}a_{\ell n}^t(\matS\matS^T)_{nn}\\
     & \geq & a_{ij}^t(\matW^T\matW)_{ii}(\matS\matS^T)_{jj}
 \end{eqnarray}
 Which when rearranged proves Equation~\ref{midproof2}.
\end{proof}

\subsection{Proof of Lemma~\ref{auxS}}
\begin{proof}
 It is clear that $G(s,s) = F_{s_{ij}}(s)$. According to the definition of an auxiliary function, it only needs to be shown that $G(s,s_{ij}^t)\geq F_{s_{ij}}(s)$. This can be done using the Taylor series expansion of $F_{s_{ij}}$:
 $$F_{s_{ij}}(s) = F_{s_{ij}}(s_{ij}^t)+F_{s_{ij}}'(s-s_{ij}^t)+\frac{1}{2}F_{s_{ij}}''(s-s_{ij}^t)^2$$
 with $F_{s_{ij}}''$ being the second order derivative of $F_{s_{ij}}$. We can show that
 $$F_{s_{ij}} = (\frac{\partial D}{\partial S})_{ij} = (-2\matA^T\matW^T
 \matX+ 2\matA^T\matW^T\matW\matA\matS)_{ij}$$
 $$F_{s_{ij}}'' = (\frac{\partial^2 D}{\partial \matS^2})_{ij} = 2(\matA^T\matW^T\matW\matA\matS)_{ii}$$
 It is now equivalent to prove
 \begin{equation}\label{midproof3}
 \frac{(\matA^T\matW^T\matW\matA\matS)_{ij}}{s_{ij}^t} \geq \frac{1}{2}F_{ij}'' = (\matA^T\matW^T\matW\matA\matS)_{ii}
 \end{equation}
 Now we have
 \begin{eqnarray}
     ((\matA^T\matW^T\matW\matA\matS)_{ij} & = & \sum_{\ell=1}^q (\matA^T\matW^T\matW\matA)_{i\ell}(\matS)_{\ell j}\\
     & \geq & (\matA^T\matW^T\matW\matA\matS)_{ii}s_{ij}^t\\
     & = & s_{ij}^t(\matA^T\matW^T\matW\matA\matS)_{ii}
 \end{eqnarray}
 which when rearranged proves Equation~\ref{midproof3}.
\end{proof}

\end{appendices}

\end{document}